# Frequency and Polarization Dependence of Thermal Coupling between Carbon Nanotubes and $SiO_2$

Zhun-Yong Ong[1,2] and Eric Pop[1,3,4,*]

[1]*Micro and Nanotechnology Lab, Univ. Illinois at Urbana-Champaign, Urbana IL 61801, U.S.A.*
[2]*Dept. of Physics, Univ. Illinois at Urbana-Champaign, Urbana IL 61801, USA*
[3]*Dept. of Electrical & Computer Engineering, Univ. Illinois at Urbana-Champaign, Urbana IL 61801, USA*
[4]*Beckman Institute, Univ. Illinois at Urbana-Champaign, Urbana IL 61801, USA*

We study heat dissipation from a (10,10) CNT to a $SiO_2$ substrate using equilibrium and non-equilibrium classical molecular dynamics. The CNT-substrate thermal boundary conductance (TBC) is computed both from the relaxation time of the CNT-substrate temperature difference, and from the time autocorrelation function of the interfacial heat flux at equilibrium (Green-Kubo relation). The power spectrum of interfacial heat flux fluctuation and the time evolution of the internal CNT energy distribution suggest that: 1) thermal coupling is dominated by long wavelength phonons between 0-10 THz, 2) high frequency (40-57 THz) CNT phonon modes are strongly coupled to sub-40 THz CNT phonon modes, and 3) inelastic scattering between the CNT phonons and substrate phonons contributes to interfacial thermal transport. We also find that the low frequency longitudinal acoustic (LA) and twisting acoustic (TA) modes do not transfer energy to the substrate as efficiently as the low frequency transverse optical (TO) mode.

*Contact: epop@illinois.edu

## I. INTRODUCTION

Energy coupling and transmission at atomic length scales is an important topic in the research on chemical reactions, molecular electronics and carbon nanotubes.[1,2] In particular, the problem of thermal coupling between carbon nanotubes (CNTs) and their environment is of great interest given the number of potential nano and microscale applications. For instance, the thermal stability of CNTs becomes important as power densities in nanoscale electronics increase,[3] or when CNT-based materials are considered for thermal management applications.[4,5]

Relevant to such scenarios is the thermal boundary conductance (TBC) between CNTs and their environment, which depends on the energy relaxation of the CNT vibrational modes. In particular, knowing the frequency or wavelength dependence of the energy relaxation channels can help tailor heat dissipation from CNTs. Previous studies[6-9] have used molecular dynamics (MD) simulations to investigate such atomistic details of energy coupling from CNTs. For instance, Carlborg et al[8] provided evidence of resonant coupling between low frequency phonon modes of a CNT and a surrounding argon matrix. Shenogin[7] found that the thermal coupling between the bending modes of the CNT and surrounding octane liquid increases with phonon wavelength.

In this study we investigate the phonon frequency dependence of CNT-$SiO_2$ energy dissipation, a system representing the majority of nanoelectronic CNT applications. This is an important follow-up to our previous work,[9] where we studied the dependence of this thermal coupling on temperature and the CNT-substrate interaction strength. We previously found that the TBC per unit CNT length is proportional to temperature, to the CNT-substrate (van der Waals) interaction strength, to CNT diameter[9] and adversely affected by substrate roughness.[10]

By contrast, the current work focuses on the frequency, wavelength and polarization dependence of thermal coupling at the CNT-$SiO_2$ interface. Here we also calculate the TBC using a new equilibrium MD (Green-Kubo) method and find good agreement with the TBC from the non-equilibrium MD method previously used.[9] From the spectral analysis of interfacial heat flux fluctuations, we find that coupling between the CNT and $SiO_2$ is stronger for low (0-10 THz) than high (10-57 THz) frequency phonons, with a peak near 10 THz. We also demonstrate that the polarization of the phonon branches affects their thermal coupling to the substrate, with the strongest component coming from the low frequency transverse optical (TO) mode. The results of this work shed fundamental insight into the atomistic energy coupling of CNTs to relevant dielectrics, and pave the way to optimized energy dissipation in CNT devices and circuits. The techniques used here can also be applied to other studies of interfacial thermal transport in nano-structures using MD simulations.

## II. SIMULATION SETUP

We use a modified version of the computational package LAMMPS (Large-scale Atomic/Molecular Massively Parallel Simulator)[9,11] for all simulations. To model the C-C interatomic interaction in the CNT we use the Adaptive Intermolecular Reactive Empirical Bond Order (AIREBO) potential.[12] To model the Si-Si, Si-O and O-O interatomic interactions in the amorphous $SiO_2$ substrate, we use the Munetoh parameterization[13] of the Tersoff potential.[14] The interatomic interactions between the CNT and the substrate atoms are modeled as van der Waals (vdW) forces using a Lennard-Jones (LJ) potential with the default parameters given in our previous work ($\chi = 1$ there).[9]

The final structure used here is a single, 98.4 Å long (10,10) CNT shown in Fig. 1. In the direction normal to the interface, the thickness of the $SiO_2$ substrate is approximately $h \approx 1.37$ nm, comparable to the height of the CNT. We have kept the system compact in this direction to avoid significant spatial variation in temperature within the $SiO_2$. Given a thermal diffusivity $\alpha \approx 10^{-6}$ m$^2$s$^{-1}$, the diffusion time in the $SiO_2$ is approximately $t_d \sim h^2/\alpha \sim 2$ ps, which is much shorter than the time scale of the thermal relaxation of the CNT, nearly 100 ps in Fig. 2a. This ensures relative temperature uniformity in the $SiO_2$ during the thermal relaxation of the CNT. In addition, the temperature decay is always monitored between the CNT and the $SiO_2$ ($\Delta T = T_{\mathrm{CNT}} - T_{\mathrm{SiO2}}$), such that any thermal build-up in the latter is automatically taken into account.

## III. CHARACTERIZATION OF THERMAL BOUNDARY CONDUCTANCE

In this study we employ two methods of characterizing the CNT-$SiO_2$ thermal boundary conductance (TBC). The first is the transient relaxation method and the second is the equilibrium Green-Kubo method, as detailed below.

### A. Relaxation Method

To simulate interfacial heat transfer, we set up an initial temperature difference $\Delta T$ between the CNT and $SiO_2$ substrate atoms, with the CNT at the higher temperature. In the absence of any coupling of the combined CNT-$SiO_2$ system to an external heat reservoir, $\Delta T$ decays expo-nentially with a single relaxation time $\tau$ i.e. $\Delta T(t) = \Delta T(0)\exp(-t/\tau)$, as shown in Fig. 2a. Given that the CNT-substrate thermal resistance (the inverse of the TBC) is much higher than the internal thermal resistance of the CNT,[15] we can apply a lumped capacitance method as in previous studies.[6,7,9] The TBC per unit length is given by $g = C_{CNT}/\tau$ where $C_{CNT} = 6.73\times10^{-12}$ JK$^{-1}$m$^{-1}$ is the *classical* heat capacity per unit length of the (10,10) CNT.

In our simulation, we set the substrate temperature to 300 K, and the initial temperature difference $\Delta T = 200$ K.[16] This is achieved by first equilibrating the atoms of the CNT and the substrate at 300 K for 100 ps. The temperature is kept at 300 K using velocity rescaling at intervals of 400 steps with a time step of 0.25 fs. Afterwards, the velocity rescaling algorithm is switched off and the system is allowed to equilibrate. To produce the temperature difference between the CNT and substrate, we again apply the velocity rescaling algorithm to the substrate atoms at 300 K and to the CNT atoms at 500 K for 10 ps.

To simulate the heat transfer process, the velocity rescaling is switched off and the system is allowed to relax without any active thermostatting. We record the decay of the temperature difference between the CNT and the substrate. Because the decay of $\Delta T$ is noisy,[16] it is necessary to average over multiple runs (twenty in this work) to obtain the exponential decay behavior shown in Fig. 2a. Here, we determine the relaxation time $\tau \approx 84$ ps and the extracted TBC per unit length for the (10,10) CNT is $g \approx 0.08$ WK$^{-1}$m$^{-1}$, consistent with our previous study.[9]

**B. Equilibrium Green-Kubo method**

At steady state in the linear regime, the CNT-substrate interfacial energy flux $Q$ is proportional to the temperature difference $\Delta T$, and is given by $Q = -g\Delta T$, where $g$ is the CNT-substrate TBC. To compute the instantaneous value of this energy flux, we use the formula

$$Q = -\frac{1}{2}\sum_{i\in \mathrm{CNT}}\sum_{j\in \mathrm{SiO_2}} \vec{f}_{ij}\bullet\left(\vec{v}_i + \vec{v}_j\right) \quad (1)$$

where $f_{ij}$ is the interatomic force between the $i$-th and $j$-th atoms and $v_i$ is the velocity of the $i$-th atom. The formula in Eq. (1) is derived in the Appendix, and corresponds to the rate of change of energy of the CNT coupled mechanically only to the substrate. More generally, this can be used to compute the energy flux between any two groups of atoms. For a micro-canonical (NVE) en-semble comprising of a CNT on a $SiO_2$ substrate, the vibrational energy and interfacial heat flux of the CNT are expected to fluctuate around their mean values.

Given that $g$ is a linear transport coefficient, it can be written at equilibrium (subscript "0") in terms of the time auto-correlation function of $Q$ using the Green-Kubo relation[17,18]

$$g_0 = \frac{1}{Lk_B T^2} \int_0^\infty dt \left\langle Q(t) Q(0) \right\rangle \tag{2}$$

where $k_B$, $T$ and $L$ are the Boltzmann constant, the CNT temperature and the CNT length respectively. To determine $g_0$, we use the same simulation setup as before in the relaxation method and thermostat the system by velocity rescaling at 300 K for 2 ns. The system is then allowed to equilibrate as an NVE ensemble for another 2 ns. We run the equilibrium simulation for 4 ns and $Q$ is computed numerically and then recorded at intervals of 1 fs.

The time auto-correlation function of $Q$ decays rapidly to zero within the first 0.5 ps. The time integral of the auto-correlation of $Q$ is shown in Fig. 2b. The numerical value of the time integral is taken to be the average from $t$ = 0.5 to 2 ns in Fig. 2b and it is 846 ± 134 $fW^2s$. On applying Eq. (2), the TBC is found to be $g_0 = 0.069 \pm 0.011$ $WK^{-1}m^{-1}$, a value in relative agreement with that obtained from the relaxation simulations. The agreement between the relaxation method and the Green-Kubo method indicates that the system is in the linear response regime. It also confirms the validity of Eq. (2) and suggests that we can compute the TBC using the Green-Kubo method alone, in itself an interesting and new result. We note that the TBC value from the relaxation method is slightly higher than that from the Green-Kubo method because the $SiO_2$ substrate heats up, as can been seen in Fig. 2a, as kinetic energy is transferred across the interface. As a result of the thermal build up in the substrate, its average temperature increases by 50 K after 160 ps. Hence, the TBC value computed from the relaxation method is one averaged over the temperature range between 300 and 350 K. Given that TBC has been found to increase with temperature,[9] we attribute the slight difference in TBC values to the heating of the $SiO_2$.

## IV. SPECTRAL ANALYSIS OF CNT-SUBSTRATE THERMAL COUPLING

We now move on to analyze the spectrum of the interfacial heat flux which gives us its phonon frequency and wavelength dependence. In addition, we study the time dependence of the CNT phonon energy distribution during the thermal relaxation process, also relating it to the interfacial heat flux spectrum.

## A. Interfacial Heat Flux Spectrum

We take the Fourier transform of the heat flux $Q$ and plot its power spectrum in Fig. 2c. Typically, in linear response theory,[19] the zero frequency limit of the spectrum is proportional to the static thermal conductance of the interface $g$, and a steady temperature discontinuity at the interface $\Delta T$. The non-zero frequency component is proportional to the Fourier transform of the response function to an oscillating *thermodynamic* force which, in this case, would be a *frequency-dependent* temperature discontinuity $\Delta T(\nu)$. Such a frequency-dependent temperature discontinuity can arise from a *coherent* non-equilibrium phonon mode at frequency $\nu$ in the CNT which would exert a driving *mechanical* force on the phonon modes in the $SiO_2$ substrate and vice versa. In the long time limit at steady state, coherence is lost and $Q$ only depends on the time-average temperature discontinuity. Therefore, we interpret the $Q(\nu)$ spectral component as the response to a *coherent* non-equilibrium phonon mode on either side of the interface. The spectrum gives us a physical picture of the dissipation of phonon energy across the CNT-substrate interface. If the spectral component is large, it suggests that the non-equilibrium CNT phonon at the corresponding frequency is more easily dissipated into the substrate.

We observe that at higher frequencies ($\nu > 10$ THz), the power spectrum scales approximately as $\nu^{-2}$. This is expected because the auto-correlation of $Q(t)$ decays exponentially, which implies that the square of its Fourier transform will behave as $\nu^{-2}$ asymptotically. However, we observe that at lower frequencies ($\nu < 10$ THz), the spectral components scale approximately as $\nu^{1/2}$ and are much larger than the higher frequency components. We have computed the same spectra for different temperatures and different CNT-substrate interaction strengths and found the shape of the spectrum to be generally the same with the dominant components between 0–10 THz. Importantly, this suggests that the interfacial thermal transport between the CNT and the $SiO_2$ substrate is dominated by low frequency phonons between 0–10 THz, and that this dominance is not affected by temperature or the CNT-substrate interaction strength.

The inset of Fig. 2c shows the linear plot of the spectrum. We observe that the spectral weight of the components from 40–57 THz is diminished compared to the spectral weight of the components from 0–40 THz, and above 57 THz the components are negligible. From 0–40 THz, the spectral components are large because, in accordance of the diffuse mismatch model,[20] the overlap in the phonon density of states (DOS) allows for transmission of phonons across the interface through elastic scattering. On the other hand, the relative weight of frequencies from 40–57 THz is diminished due to the absence of overlapping phonon modes between the CNT and $SiO_2$ in this frequency range, as can be seen in Fig 2d. In other words, CNT phonons from 40-57 THz cannot be elastically scattered into the substrate and must undergo inelastic scattering by the interface in order to be transmitted. Nevertheless, they still contribute to the TBC as can be seen in Fig. 2c. This supports the finding in our earlier work[9] that inelastic scattering at the interface plays an important role in the temperature dependence of the TBC. The spectral weight above 57 THz is negligible because such high frequencies are not supported in either the CNT or the $SiO_2$ substrate.

To provide additional insight into how inelastic scattering at the interface is reflected in the spectrum of the interfacial heat flux, we use the simple model of vibrational energy transfer by Moritsugu et al.[21] This consists of an oscillator 1 of frequency $\omega_1$ coupled to an oscillator 2 of frequency $\omega_2$, with $\omega_2 > \omega_1$, via a third order anharmonic term. The Lagrangian of this system can be written as

$$L = \frac{1}{2}\left(\dot{q}_1^2 + \dot{q}_2^2\right) - \frac{1}{2}\left(\omega_1^2 q_1^2 + \omega_2^2 q_2^2\right) - \alpha q_1^2 q_2 \tag{3}$$

where $\alpha$ is the coupling coefficient and $q_i$ is the normal coordinate of oscillator $i$. The anharmonic coupling term is quadratic in $q_1$ and linear in $q_2$ to model the inelastic scattering of one $\omega_2$ phonon to two $\omega_1$ phonons (a bilinear coupling term would simply lead to resonant energy transfer between the two oscillators). The harmonic energy of oscillator 2 is, to linear order in $\alpha$,

$$E_2 = \frac{1}{2}A_2^2\omega_2^2 - \frac{\alpha A_1^2 A_2}{4}\left[\frac{\omega_2}{2\omega_1+\omega_2}\cos\left(2\tau_1+\tau_2\right) - \frac{\omega_1}{2\omega_1-\omega_2}\cos\left(2\tau_1-\tau_2\right) + 2\cos\left(\tau_2\right)\right] \tag{4}$$

where $A_i$ is the amplitude and $\tau_i = \Omega_i t + \theta_i$ with $\Omega_i = \omega_i + O(\alpha)$ being the modulated frequency and $\theta_i$ being the initial phase of oscillator $i$. As oscillator 2 is coupled only to oscillator 1, its rate of change of energy is equal to the energy flux between the two oscillators $Q_{21}$ and is given by

$$Q_{21} = \frac{dE_2}{dt} = \frac{\alpha A_1^2 A_2 \omega_2}{4}\left[\sin\left(2\tau_1+\tau_2\right) - \sin\left(2\tau_1-\tau_2\right) + 2\sin\left(\tau_2\right)\right] \tag{5}$$

The last term in Eq. (5) is sinusoidal with frequency $\omega_2$. Thus, the Fourier transform of the auto-correlation of $Q_{21}$ would give us a component of frequency $\omega_2$ in the power spectrum. If oscillator 1 were a non-overlapping phonon mode in the CNT and oscillator 2 a phonon mode in the $SiO_2$ such that $\omega_2 \sim 2\omega_1$, then the direct anharmonic coupling between them would contribute the component of the interfacial flux spectrum at $\omega_2$. Classically, this corresponds to the coupling of the first harmonic of oscillator 1 to the second harmonic of oscillator 2.

**B. Time-dependent CNT Phonon Energy Distribution**

To gain further insight into the role of low frequency phonons in interfacial transport, we track the time evolution of their energy distribution during the relaxation process. As mentioned above, the power spectrum suggests the interfacial thermal transport is dominated by low frequency phonons from 0–10 THz. If this were so, the phonon energy distribution should change during the relaxation simulations since low frequency phonons would cool more rapidly.

The phonon energy distribution can be obtained by analyzing the energy spectrum of the velocity fluctuations of the atoms in the CNT. The velocities of the CNT atoms are saved at intervals of 5 fs over 200 ps. As the energy relaxation is a non-stationary process, we employ a windowing Fourier transform technique to compute the moving power spectrum of the CNT, to obtain both time and frequency information. The translational and rotational symmetries of the armchair CNT means that we can also perform the Fourier transform with respect to its axial and angular positions to gain additional information on wave vector and angular symmetry dependence. In each window, we calculate the CNT power spectrum

$$\Theta(t,\nu,k,\alpha) = \frac{m}{2}\sum_{b=1}^{p}\sum_{\zeta=x,y,z}\left|\frac{1}{NM\tau}\sum_{n=0}^{N-1}\sum_{\alpha=0}^{M-1}\left(\exp\left[2\pi i\left(\frac{kn}{N}+\frac{\alpha\theta}{M}\right)\right]\int_{t}^{t+\tau} v_{b,\zeta}(n,t')e^{-2\pi i\nu t'}dt'\right)\right|^2 \quad (6)$$

where $m$ is the mass of the C atom, $t$ the time of the window, $\nu$ is the frequency, $k$ is the wave vector (from 0 to half the total number of axial unit cells), $\alpha$ is the angular dependence around the tube (from 0 to half the total number of repeat units circumferentially), $p$ is the number of basis atoms in each unit cell ($p$ = 4 for armchair CNTs), $N$ is the total number of axial unit cells, $M$ is the total number of circumferential repeat units, and $\tau$ = 5 ps the window period. We choose the z-axis to be parallel to the CNT. The raw spectral density obtained from Eq. (6) estimates the energy distribution in ($\nu,k,\alpha$) space. The transverse power spectrum can be obtained from Eq. (6)

by summing over only the $x$ and $y$ components of the velocities instead of all three components. Similarly, the longitudinal spectrum can be obtained by summing over only the $z$ component of the atomic velocities.

At equilibrium, the CNT power spectrum is time-independent and given as

$$\Theta_{Eq}\left(\nu,k,\alpha\right)=k_B T_{Eq}\rho\left(\nu,k,\alpha\right) \tag{7}$$

where $\rho(\nu,k,\alpha)$ is the normalized *spectral density of states*, $k_B$ the Boltzmann constant and $T_{Eq}$ the equilibrium temperature. At non-equilibrium, we can rewrite Eq. (7) as

$$\Theta_{Neq}\left(t,\nu,k,\alpha\right)=k_B T_{Sp}\left(t,\nu,k,\alpha\right)\rho\left(\nu,k,\alpha\right) \tag{8}$$

where $T_{Sp}(t,\nu,k,\alpha)$ is the spectral temperature dependent on frequency, wave vector and circumferential angle. This is not the temperature in the strict thermodynamic sense, but a measure of the energy distribution in $(\nu,k,\alpha)$ space at non-equilibrium.

### C. 2D Power Spectrum of CNT at Equilibrium

To obtain the transverse, longitudinal and overall power density maps shown in Figs. 3a, b and c, we sum over circumferential angle $\alpha$ in Eq. (7) and plot the resultant 2D power spectrum of the CNT in $(\nu,k)$ space. If we compare Figs. 3a and 3b at small wave vector and low frequency, we note that most of the energy distribution is for transversely polarized phonon branches. We also observe that the transverse power spectrum has many more low frequency optical phonon branches from 0–10 THz than the longitudinal power spectrum. This lends additional credence to our earlier observation that the heat flux fluctuations from 0–10 THz dominate the power spectrum in Fig. 2c.

It is clear that the phonon dispersion relation can also be obtained from the power spectrum.[22] Theoretically, there should be 120 branches (degenerate and non-degenerate) in the phonon spectrum of a (10,10) CNT, and most of the individual phonon branches can be resolved in the 2D power spectrum. Taking advantage of the $D_{10}$ symmetry of the (10,10) CNT, the 2D power spectrum in Fig. 3c can be further decomposed into $\alpha$-dependent sub-spectra. We plot $\Theta_{Neq}(\nu,k,\alpha)$ from Eq. (7) for individual values of $\alpha$ (0 to 5) in Fig. 4a and obtain the individual 2D angular-dependent phonon sub-spectrum. Rotational symmetry dictates there ought to be 12 branches for each sub-spectrum and they can be clearly seen in each of the plots. The longitudin-al acoustic (LA) and twisting acoustic (TW) modes can be distinguished in the small $k$, low $\nu$ region of the $\alpha = 0$ sub-spectrum. Similarly, the flexural acoustic modes can be seen in the corresponding region of the $\alpha = 1$ sub-spectrum as predicted by Mahan and Jeon.[23] For higher angular numbers, only optical modes can be found in the small $k$, low $\nu$ region. These low frequency phonon modes can have very different interfacial transport behavior. In Fig. 4b, we compare the small $k$, low $\nu$ region of the $\alpha = 0$ and $\alpha = 5$ sub-spectra. The former has a pair of acoustic branches while the latter consists of a pair of degenerate transverse optical branches.

## D. Spectral Temperature Decays

We define the average spectral temperature of a certain region in the ($\nu,k,\alpha$) space as

$$T_{Sp}\left(t,\nu_{\min},\nu_{\max},k_{\min},k_{\max},\alpha_{\min},\alpha_{\max}\right)=\frac{\sum_{k_{\min}}^{k_{\max}}\sum_{\alpha_{\min}}^{\alpha_{\max}}\int_{\nu_{\min}}^{\nu_{\max}}\Theta_{Neq}\left(t,\nu',k,\alpha\right)d\nu'}{k_B\sum_{k_{\min}}^{k_{\max}}\sum_{\alpha_{\min}}^{\alpha_{\max}}\int_{\nu_{\min}}^{\nu_{\max}}\rho\left(\nu',k,\alpha\right)d\nu'} \tag{9}$$

This definition is a weighted average, unlike the unweighted temperature previously employed by Carlborg et al.[8] When integrated over the entire ($\nu,k,\alpha$) space, the expression above leads to the average temperature of the entire CNT.

It has been previously suggested that long-wavelength and low-frequency modes are key to the heat flow across the interface.[6,7] To further our understanding of the wavelength and frequency dependence, we compute the spectral temperatures for the frequency and normalized wave vector ranges in Table 1. $T_{Sp}^{(I)}$, $T_{Sp}^{(II)}$ and $T_{Sp}^{(II)}$ are the average spectral temperatures of regions I (small $k$, low $\nu$), II (large $k$, low $\nu$) and III (small $k$, high $\nu$) of the spectrum, as shown in Fig. 3c. $T_{Sp}^{(IV)}$ is the average spectral temperature for the non-overlapping region. $T_{Sp}^{(V)}$, $T_{Sp}^{(VI)}$ and $T_{Sp}^{(VII)}$ are the average spectral temperatures for different values of $\alpha$. $T_{Sp}^{(VIII)}$ is the spectral temperature of the LA and TW phonon branch in the low-frequency, long-wavelength regime of the $\alpha = 0$ sub-spectrum while $T_{Sp}^{(IX)}$ is the spectral temperature of the TO phonon branch in the small $k$, low $\nu$ regime of the $\alpha = 5$ sub-spectrum.

### D.1. Frequency and wave vector dependence

To gain insight into this heat transfer mechanism, we compare the spectral temperature decay for $T_{Sp}^{(I)}$, $T_{Sp}^{(II)}$ and $T_{Sp}^{(III)}$ to the decay for $T_{CNT}$ by computing $\Delta T_{Sp}^{(I)}= T_{Sp}^{(I)}-T_{Sub}$, $\Delta T_{Sp}^{(II)}=$ $T_{Sp}^{(II)}$-$T_{Sub}$ and $\Delta T_{Sp}^{(III)}$= $T_{Sp}^{(III)}$-$T_{Sub}$ where $T_{Sub}$ is the temperature of the substrate. We compare their time evolution with that of the overall temperature decay $\Delta T$.

The time evolutions for $\Delta T_{Sp}^{(I)}$ and $\Delta T_{Sp}^{(II)}$ are shown in Figs. 5a and 5b. We observe that the time evolution of $\Delta T_{Sp}^{(I)}$ (small $k$, low $\nu$) displays two stages: a rapid initial followed by a slower exponential decay with a characteristic time of ~84 ps which is about the same as the decay time for $\Delta T$. This suggests that there is a rapid initial energy transfer from the CNT to the substrate in this regime. Another possibility is that some of the small $k$, low $\nu$ modes relax much more rapidly whiles the remaining ones relax at the same rate as the rest of the CNT. Intuitively, this makes sense because low frequency, long wavelength modes experience more deformation than higher frequency modes when the CNT is in contact with the substrate.

On other hand, $\Delta T_{Sp}^{(II)}$ (large $k$, low $\nu$) and $\Delta T_{Sp}^{(III)}$ (small $k$, high $\nu$) decay at the same rate as $\Delta T$ and they can be fitted with a single exponential decay. It has been suggested that the primary heat transfer mechanism between the CNT and its surrounding medium is the coupling of the low frequency long wavelength modes in the CNT to the vibrational modes of the surrounding medium[6]. The absence of any rapid decay in $\Delta T_{Sp}^{(III)}$ (small $k$, high $\nu$) excludes the higher frequency, long wavelength modes from playing a significant role in the CNT-substrate energy transfer process. Similarly, the normal decay of $\Delta T_{Sp}^{(II)}$ (large $k$, low $\nu$) supports the idea that only short wavelength modes are weakly coupled to the substrate phonons and do not contribute significantly to the CNT-substrate energy transfer process. Our data support the hypothesis that low frequency and long wavelength phonons are the primary modes of energy dissipation from the CNT into the surrounding medium.

We offer the following qualitative explanation. The higher frequency modes generally involve significant bond stretching and compression *within* the unit cell and any mechanical coupling to the substrate is unlikely to lead to significant deformation and hence, thermal coupling. On the other hand, shorter wavelength phonon modes also involve significant bond stretching and compression *between* unit cells along the tube axis. Thus, the mechanical coupling to the substrate would also be relatively limited.

### D.2. Non-overlapping region (over-40 THz)

We now examine the temperature decay $\Delta T_{\mathrm{Sp}}^{(\mathrm{IV})}$ ($\nu > 40$ THz) for the high frequency phonon modes of the CNT which do not have counterparts in the $SiO_2$. If the interfacial thermal transport is primarily modulated by phonons in the overlap region or if the energy transfer rate between the overlapping and non-overlapping regions is comparable to or slower than the heat transfer rate to the substrate, then $\Delta T_{\mathrm{Sp}}^{(\mathrm{IV})}$ would decay at a slower rate than $\Delta T$. However, we find no discernible difference between the decay rate of $\Delta T_{\mathrm{Sp}}^{(\mathrm{IV})}$ and that of $\Delta T$. This suggests that vibrational energy in the non-overlapping region is dissipated, directly or indirectly, into the substrate at the same rate as the overlapping region. It also implies that the intra-nanotube energy exchange rate between the overlapping and non-overlapping regions is much higher than the energy transfer rate from the CNT into the substrate, allowing the energy distribution in the over-40 THz region to be in quasi-equilibrium with the sub-40 THz region. We conclude that there is no bottleneck in the transfer of energy from the over-40 THz CNT phonon modes to the substrate because they are directly transmitted through inelastic scattering and indirectly through intra-nanotube scattering into lower frequency, small wave vector phonons which are then transmitted into the substrate.

### D.3. Angular and polarization dependence

Here, we compute the decay times for spectral temperatures $\Delta T_{\mathrm{Sp}}^{(\mathrm{IV})}$ ($\alpha = 1, 2$), $\Delta T_{\mathrm{Sp}}^{(\mathrm{V})}$ ($\alpha = 3, 4$) and $\Delta T_{\mathrm{Sp}}^{(\mathrm{VI})}$ ($\alpha = 5, 6$) in order to determine the angular dependence, if any. However, we do not find any significant difference between their decay times and that of $\Delta T$. Their time evolutions are identical to that of $\Delta T_{\mathrm{Sp}}^{(\mathrm{II})}$. This implies that there is no explicit dependence of the CNT-substrate thermal coupling on the angular number alone.

We also plot in Figs. 5c and 5d the temperature decays for $\Delta T_{\mathrm{Sp}}^{(\mathrm{VIII})}$ (small $k$, low $\nu$ TW and LA branches) and $\Delta T_{\mathrm{Sp}}^{(\mathrm{IX})}$ (small $k$, low $\nu$ TO branches). We find that the decay rate for $\Delta T_{\mathrm{Sp}}^{(\mathrm{IX})}$ is significantly greater than that of $\Delta T_{\mathrm{Sp}}^{(\mathrm{VIII})}$. $\Delta T_{\mathrm{Sp}}^{(\mathrm{IX})}$ decays at approximately the same rate as the rest of the CNT. This implies that, despite the long wavelengths and low frequencies of the LA and TW acoustic modes, they are weakly coupled to the substrate and thermalize rapidly with the other higher-frequency phonon modes. This is not surprising, because the LA and TW phonon modes are due to in-plane displacements, and the compression or shearing motion results in weak mechanical coupling to the substrate. On the other hand, the transverse optical modes represented by $\Delta T_{Sp}^{(IX)}$ are much more strongly coupled to the substrate than they are to the rest of the CNT. This is because these modes have radial displacements, resulting in stronger mechanical and thermal coupling.

## V. CONCLUSIONS

We have examined the interfacial thermal transport between a (10,10) CNT and a $SiO_2$ substrate with equilibrium and non-equilibrium molecular dynamics simulations. We computed the thermal boundary conductance (TBC) using a Green-Kubo relation and found that it is in agreement with the value obtained from the relaxation method.[9] This demonstrates that the TBC can be computed independently using the Green-Kubo relation in equilibrium MD simulations. By a spectral analysis of the heat flux, we found that thermal coupling across the CNT-$SiO_2$ interface is most strongly mediated by low frequency, 0–10 THz phonons. We also found a small contribution to the interfacial heart flux fluctuations in the 40-57 THz region, which suggests that high frequency CNT phonons couple directly into the substrate through inelastic scattering.

Furthermore, we tracked the time evolution of the CNT phonon energy distribution during the relaxation simulations. We found that low-frequency phonon modes (0–10 THz) relax more rapidly, confirming their primary role in interfacial thermal transport. However, short wavelength phonons relax at the same rate as the rest of the CNT even at low frequencies. The overlap between the phonon spectra of the CNT and the $SiO_2$ substrate from 0-40 THz does not seem to have a significant effect on the energy relaxation of the phonon modes. The spectral temperature in the over-40 THz region of the phonon spectrum is found to relax at the same rate as the rest of the CNT. This suggests that that the intra-nanotube energy exchange rate between the overlapping and non-overlapping regions of the phonon spectrum is much higher than the energy transfer rate from the CNT into the substrate, allowing the energy distribution in the over-40 THz region to be in quasi-equilibrium with the sub-40 THz region. We also find no explicit dependence of the energy relaxation on the angular number.

Finally, we computed the energy relaxation rate of LA and TW phonon branches in the small $k$, low $\nu$ region of the phonon spectrum and compared it to that of one of the TO phonon branches. The energy transfer rate was found to be much higher for the TO modes than for the LA and TW modes, suggesting that transverse vibrations of the CNT are primarily responsible for thermal coupling in this configuration. The results of this work shed key insight into the thermal coupling of CNTs to relevant dielectrics, and pave the way to optimized energy dissipation in CNT devices and circuits. The techniques used here could also be applied to other studies of interfacial thermal transport using MD simulations.

## ACKNOWLEDGEMENT

This work has been partly supported by the Nanoelectronics Research Initiative (NRI) SWAN center, the NSF CCF 08-29907 grant, and a gift from Northrop Grumman Aerospace Systems (NGAS). The molecular images in Fig. 1 were generated using the graphics program VMD.[24] We acknowledge valuable technical discussions with Junichiro Shiomi, and computational support from Reza Toghraee and Umberto Ravaioli.

## APPENDIX: Derivation of Interfacial Heat Flux

We derive the expression of the interfacial heat flux here. For simplicity in our derivation, we only consider the 1-dimensional case and assume pairwise interactions between the atoms. Let the Hamiltonian $H$ of the system be

$$H = \sum_{i=1}^{N} \frac{p_i^2}{2m_i} + \sum_{i=1}^{N} \sum_{j=i+1}^{N} V_{ij} = \sum_{i=1}^{N} \frac{p_i^2}{2m_i} + \frac{1}{2} \sum_{i=1}^{N} \sum_{j=1}^{N} V_{ij} = \sum_{i=1}^{N} H_i \quad (10)$$

where $m_i$ and $p_i$ are the mass and momentum of the $i$-th atom and $V_{ij}$ is the interaction potential energy between the $i$-th and $j$-th atoms. The energy of the $i$-th atom is given by

$$H_i = \frac{p_i^2}{2m_i} + \frac{1}{2} \sum_{j=1}^{N} V_{ij} \quad (11)$$

For convenience, we set $V_{ii}$ the self energy of the $i$-th atom equal zero. The time derivative of $H_i$ can be obtained by taking its Poisson bracket[25] with $H$

$$\frac{dH_i}{dt} = \{H_i, H\} = \sum_{j=1}^{N} \left( \frac{\partial H_i}{\partial q_j} \frac{\partial H}{\partial p_j} - \frac{\partial H_i}{\partial p_j} \frac{\partial H}{\partial q_j} \right). \quad (12)$$

Substituting Eqs. (10) and (11) into Eq. (12) and using the relationship

$$\dot{q}_j = \frac{\partial H}{\partial p_j}, \quad (13)$$

we evaluate the expression in Eq. (13) to be

$$\frac{dH_i}{dt} = \sum_{j=1}^{N}\left[\dot{q}_j \frac{\partial H_i}{\partial q_j} - \frac{\partial H}{\partial q_j}\frac{\delta_{ij} p_j}{m_j}\right] = \sum_{j=1}^{N}\dot{q}_j \frac{\partial H_i}{\partial q_j} - \frac{p_i}{m_i}\sum_{j=1}^{N}\frac{\partial H_j}{\partial q_i} = -\sum_{j=1}^{N} Q_{ij} \tag{14}$$

where we have defined $Q_{ij}$, the energy flux from the *i*-th atom to the *j*-th atom, as

$$Q_{ij} = \dot{q}_i \frac{\partial H_j}{\partial q_i} - \dot{q}_j \frac{\partial H_i}{\partial q_j}. \tag{15}$$

Note that $Q_{ij} = -Q_{ji}$ and $Q_{ii} = 0$ as expected. The partial derivatives in Eq. (15) are evaluated below as

$$\frac{\partial H_i}{\partial q_j} = \frac{\partial}{\partial q_j}\left(\frac{1}{2}\sum_{k=1}^{N} V_{ik}\right) = \frac{1}{2}\frac{\partial V_{ij}}{\partial q_j} = \frac{1}{2} f_{ij} \tag{16}$$

where $f_{ji}$ is the force exerted by atom *i* on atom *j*. There is a factor of ½ because of the way we divide up the interaction energy between the atoms in Eq. (11). Thus, Eq. (15) becomes

$$Q_{ij} = -\frac{1}{2}\left(\dot{q}_i f_{ij} - \dot{q}_j f_{ji}\right) = -\frac{1}{2}\left(\dot{q}_i + \dot{q}_j\right) f_{ij}. \tag{17}$$

Therefore, the total energy flux from a group of atoms A into another group of atoms B is

$$Q_{\mathrm{A}\to\mathrm{B}} = \sum_{i\in\mathrm{A}}\sum_{j\in\mathrm{B}} Q_{ij} = -\frac{1}{2}\sum_{i\in\mathrm{A}}\sum_{j\in\mathrm{B}}\left(\dot{q}_i + \dot{q}_j\right) f_{ij} \tag{18}$$

In 3 dimensions, the total energy flux from A to B is

$$Q_{\mathrm{A}\to\mathrm{B}} = -\frac{1}{2}\sum_{i\in\mathrm{A}}\sum_{j\in\mathrm{B}} \vec{f}_{ij} \bullet \left(\vec{v}_i + \vec{v}_j\right). \tag{19}$$

**FIGURES:**

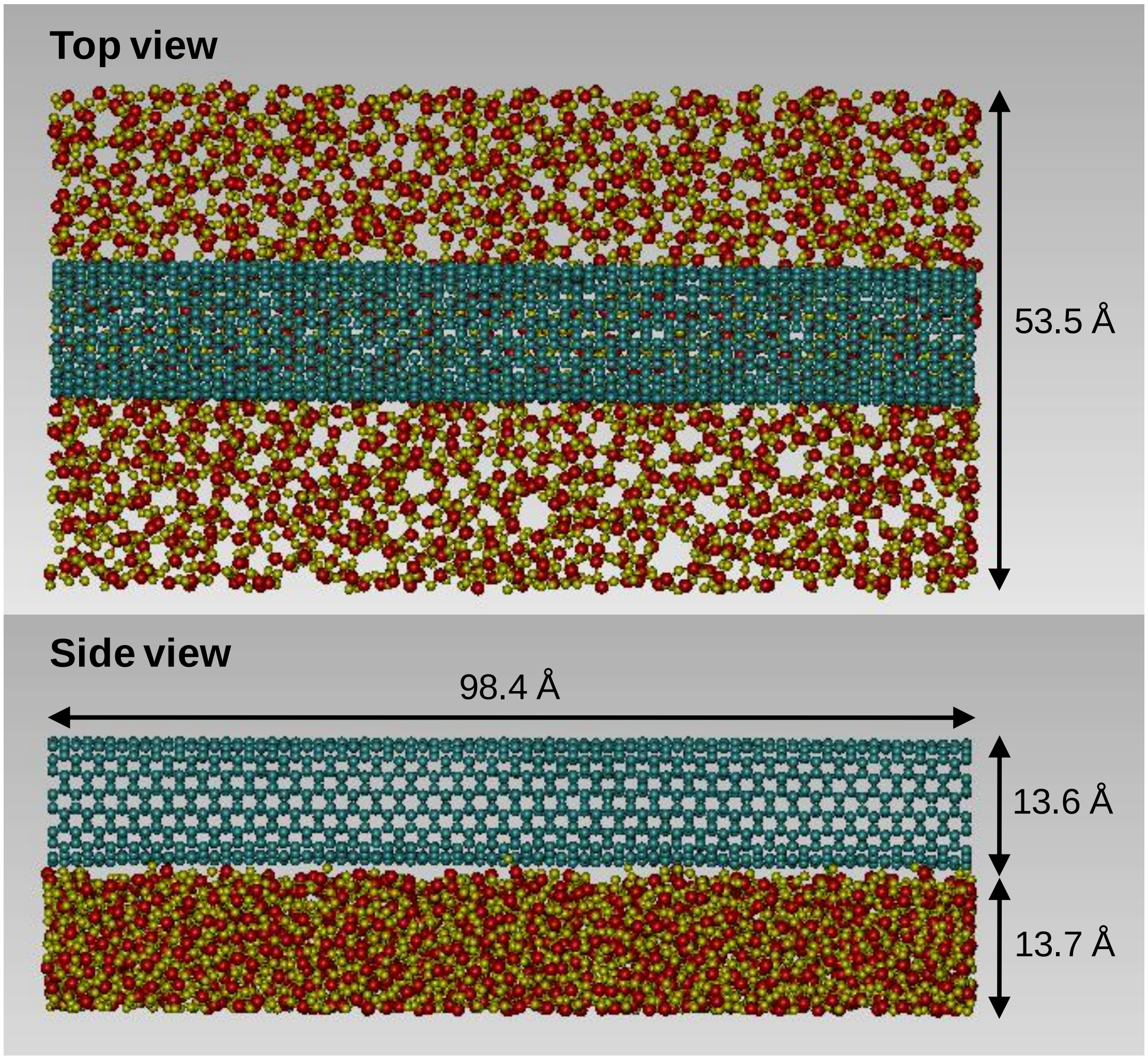

**FIG. 1.** (Color Online) Top and side view of the simulation setup for non-equilibrium energy dissipation from a (10,10) CNT to the $SiO_2$ substrate. The CNT consists of 1600 atoms, is 40 unit cells (98.4 Å) long and 13.6 Å in diameter. The amorphous $SiO_2$ substrate is produced by annealing a β-cristobalite crystal at 6000 K and 1 bar for 10 ps before slow quenching to 300 K at a rate of $10^{12}$ $Ks^{-1}$. Its final dimensions are 53.5 Å by 13.7 Å by 98.4 Å. The CNT is then placed on top of the slab. The final structure is obtained after energy minimization.

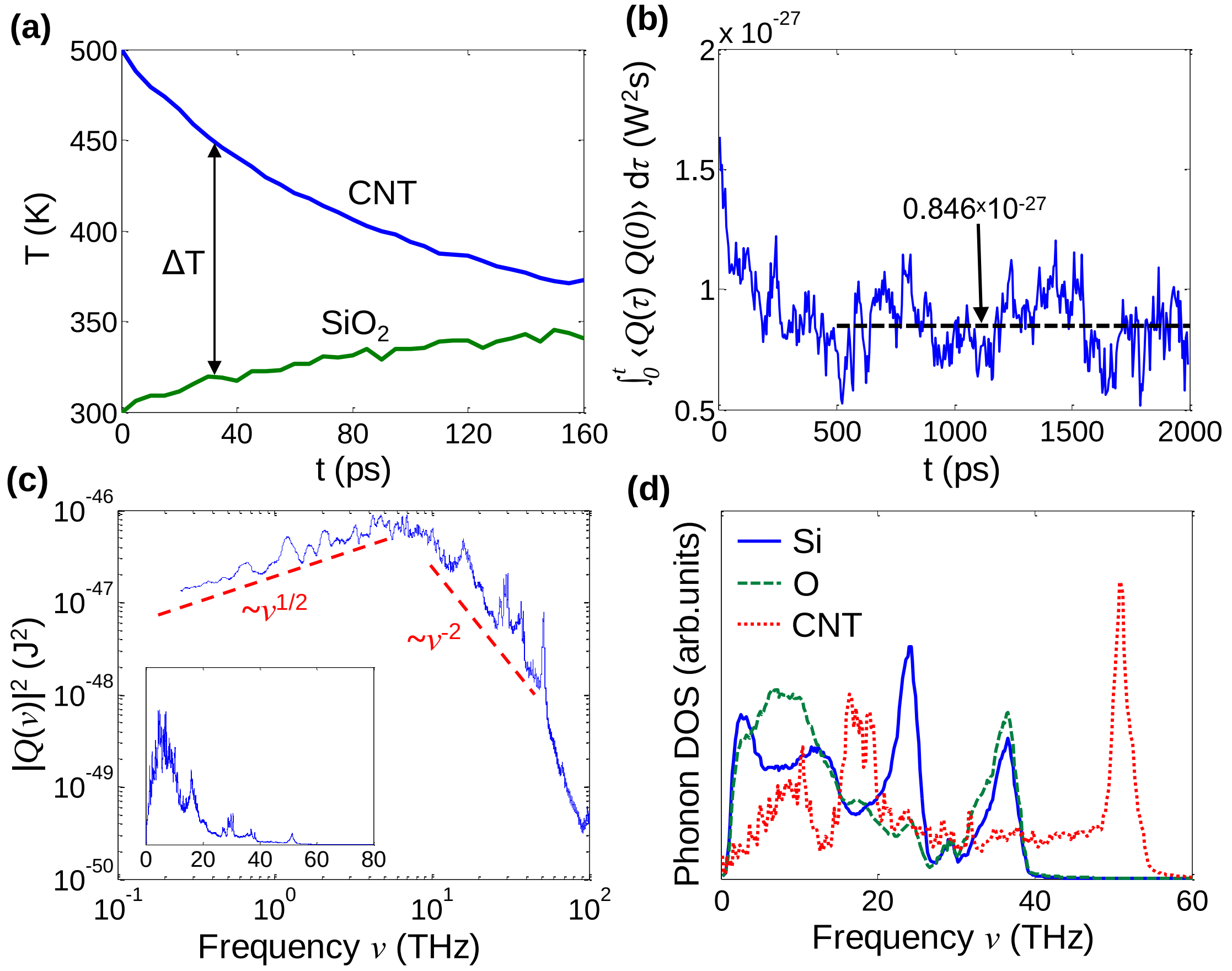

**FIG. 2.** (Color Online) **(a)** Time evolution of the CNT and $SiO_2$ temperature, showing thermal time constant ~84 ps which corresponds to $g$ ~ 0.080 WK$^{-1}$m$^{-1}$. **(b)** Time integral of the auto-correlation of $Q$ oscillates about its asymptotic value of 0.846x10$^{-27}$ W$^2$s after ~500 ps. This corresponds to $g$ ~ 0.069 ± 0.011 WK$^{-1}$m$^{-1}$. **(c)** Log-log plot of the power spectrum of $Q$ scales as $\nu^{1/2}$ from 0-10 THz, and as $\nu^{-2}$ at higher frequencies. Inset shows the same as a linear plot; both suggest the dominant contribution is between 0 -10 THz. A small component of very high frequency (>40 THz) phonons in the CNT also contribute to interfacial thermal transport via inelastic scattering, although there are no corresponding modes in the $SiO_2$ substrate. **(d)** Normalized phonon density of states (DOS) for the atoms in the $SiO_2$ substrate and the atoms in the CNT.

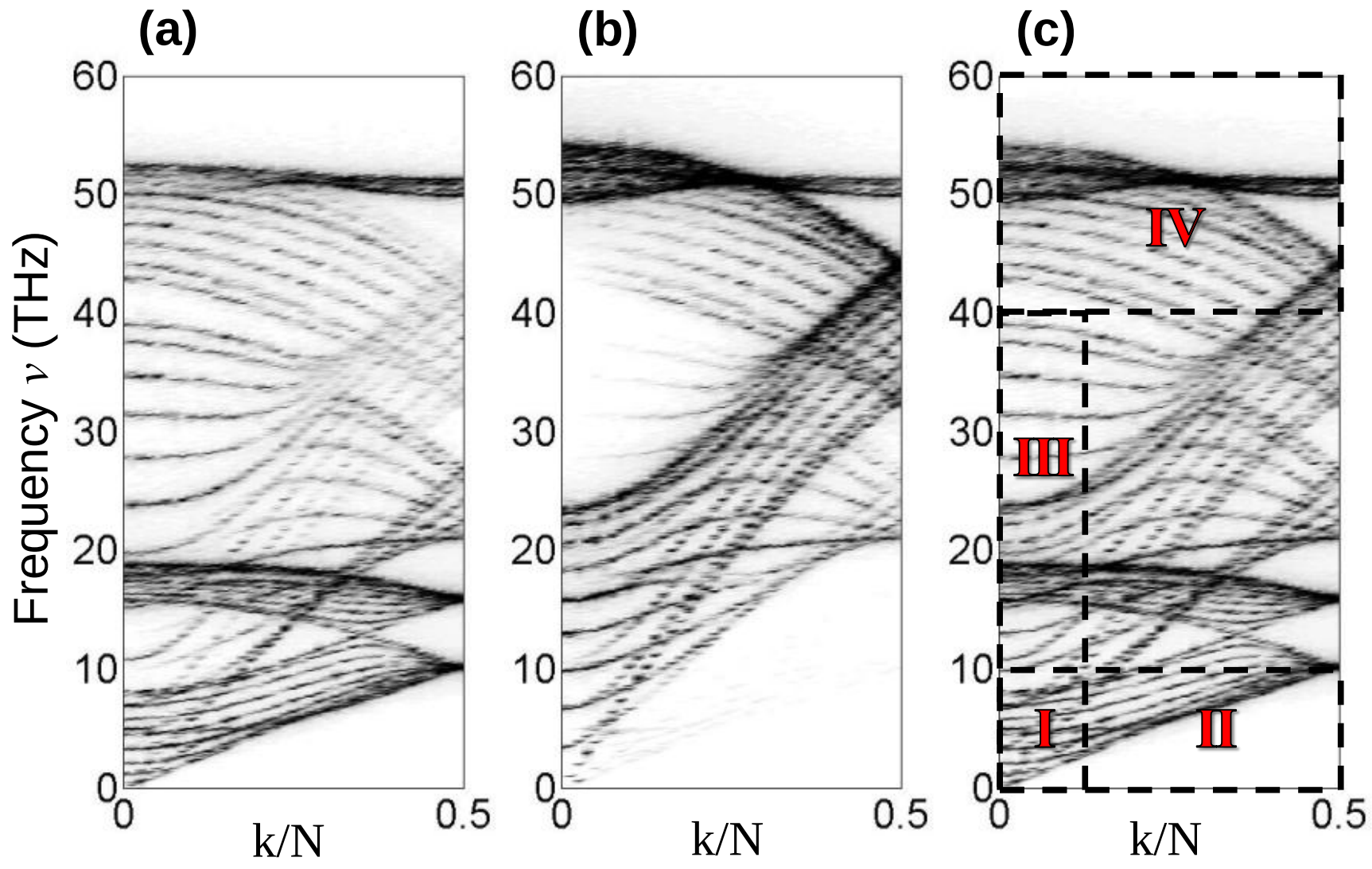

**FIG. 3.** Power spectrum of the isolated (10,10) CNT. **(a)** Transverse power spectrum. Closely spaced optical phonon branches are visible in the low-frequency part. **(b)** Longitudinal power spectrum. Unlike the transverse power spectrum, it has fewer low frequency optical phonon branches. **(c)** Overall power spectrum as the sum of the transverse and longitudinal power spectra. We divide the power spectrum into different regions (I, II, III and IV) to compute their spectral temperatures.

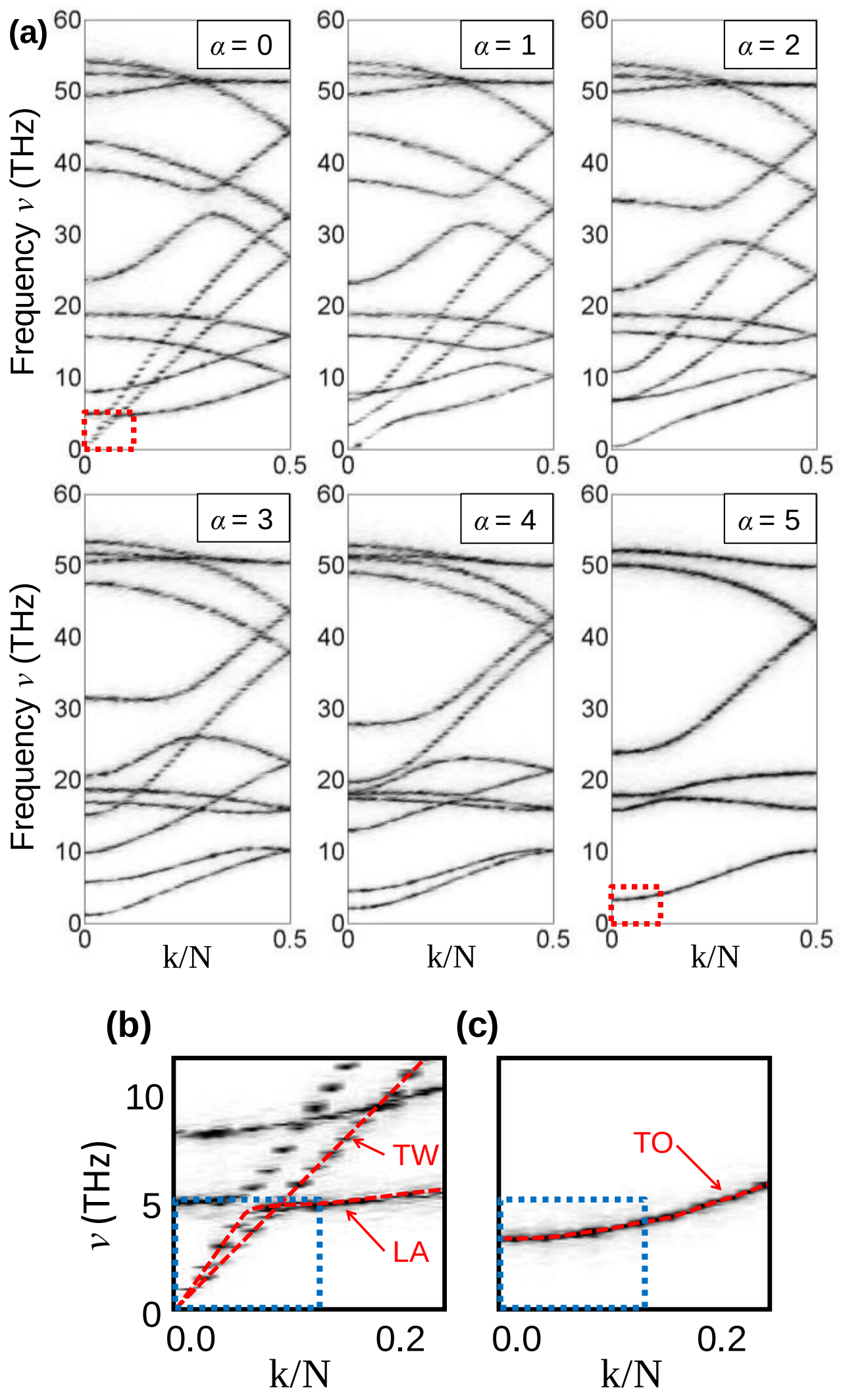

**FIG. 4.** (Color Online) **(a)** Power spectrum of an isolated (10,10) CNT for $\alpha$ = 0 (top left) to 5 (bottom right). There are 12 phonon branches for each value of $\alpha$ (doubly degenerate for $\alpha$ = 5). By decomposing the power spectrum with respect to $\alpha$, we observe the distinct phonon branches. **(b)** Detail near the origin of the $\alpha$ = 0 plot, showing longitudinal acoustic (LA) and twisting acoustic (TW) branches. **(c)** Detail near the origin of the $\alpha$ = 5 plot showing a doubly degenerate purely transverse optical (TO) branch. The regions in **(b)** and **(c)** enclosed by the blue dotted lines are used for computing the spectral temperatures $T_{\mathrm{Sp}}^{(\mathrm{VIII})}$ and $T_{\mathrm{Sp}}^{(\mathrm{IX})}$ respectively. They are also shown with dotted lines in the $\alpha$ = 0 and $\alpha$ = 5 power spectra.

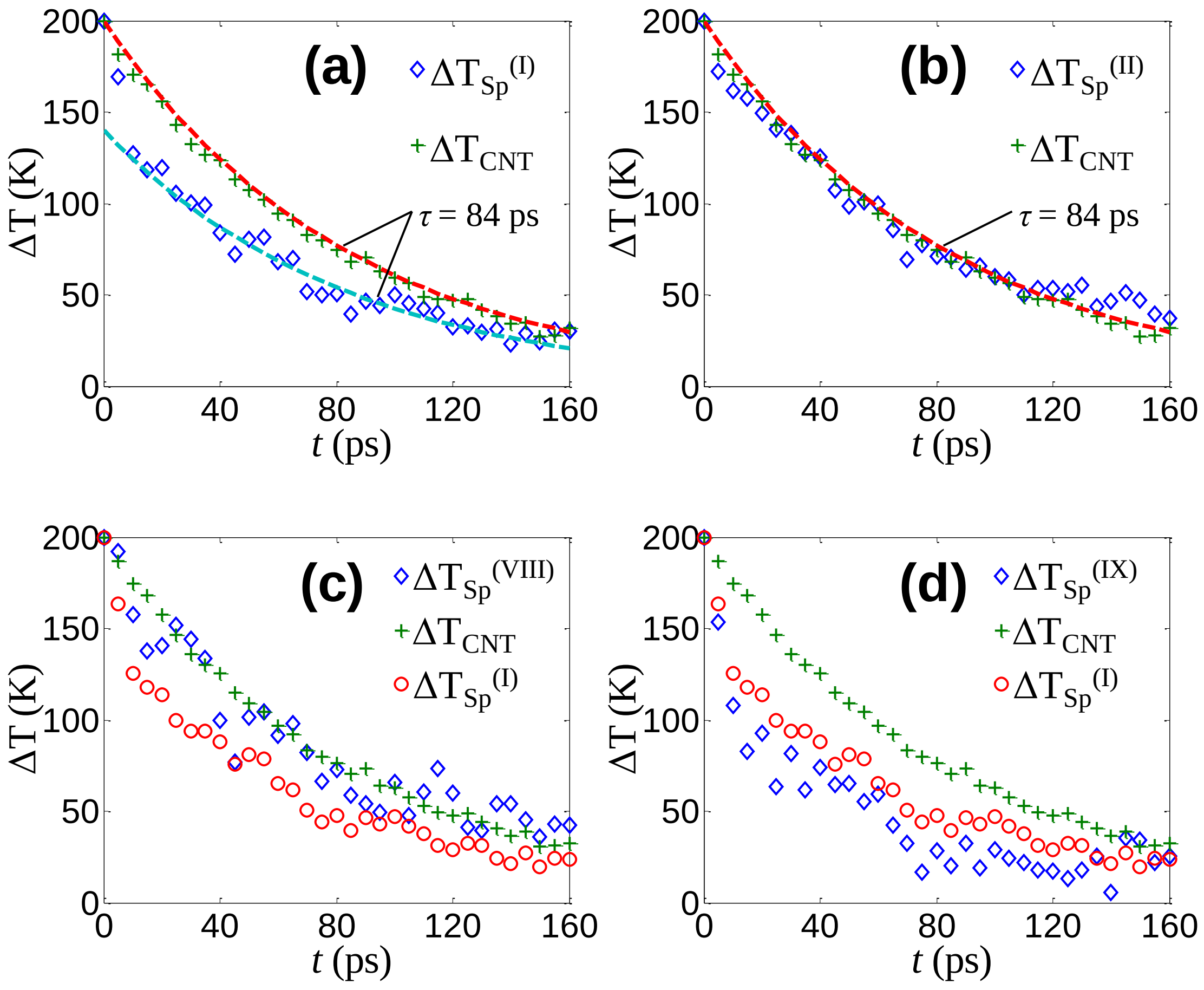

**FIG. 5.** (Color Online) Spectral temperature decays of the (10,10) CNT with the initial CNT temperature at 500 K and the $SiO_2$ substrate at 300 K. **(a)** $\Delta T_{Sp}^{(I)}$ exhibits a fast initial decay followed by a slower decay rate of $\tau \sim$ 84 ps. **(b)** Shows the spectral temperature decay of $\Delta T_{Sp}^{(II)}$. It relaxes at about the same rate as $\Delta T_{Sp}^{(II)}$ to $\Delta T_{Sp}^{(VII)}$ and the average temperature of the CNT. The relaxation behavior of $\Delta T_{Sp}^{(I)}$ suggests that small $k$, low $\nu$ phonon modes are much more strongly coupled to the substrate phonons and are the primary mechanism responsible for vibrational energy transfer to the substrate. **(c)** and **(d)** show the spectral temperature decay of $\Delta T_{Sp}^{(VIII)}$ and $\Delta T_{Sp}^{(IX)}$ compared to $\Delta T_{Sp}^{(I)}$ and $\Delta T_{CNT}$. $\Delta T_{Sp}^{(VIII)}$, which corresponds to the energy relaxation of the small $k$, low $\nu$ LA and TW modes, decays at approximately the same rate as $\Delta T_{CNT}$. On the other hand, $\Delta T_{Sp}^{(IX)}$, which corresponds to the energy relaxation of the small $k$, low $\nu$ TO mode for $\alpha$ = 5, decays even more rapidly than $\Delta T_{Sp}^{(I)}$.

**Table 1**: Frequency, angular number and wave vector ranges for the spectral temperatures.

| Case | Spectral Temperature | $\nu$ range (THz) | $k$ range | $\alpha$ range |
|---|---|---|---|---|
| 1 | $T_{\mathrm{Sp}}^{(\mathrm{I})}$ | 0 to 10 | 0 to 5 | 0 to 5 |
| 2 | $T_{\mathrm{Sp}}^{(\mathrm{II})}$ | 0 to 10 | 6 to 20 | 0 to 5 |
| 3 | $T_{\mathrm{Sp}}^{(\mathrm{III})}$ | 10 to 40 | 0 to 5 | 0 to 5 |
| 4 | $T_{\mathrm{Sp}}^{(\mathrm{IV})}$ | 40 to 60 | 0 to 20 | 0 to 5 |
| 5 | $T_{\mathrm{Sp}}^{(\mathrm{V})}$ | 0 to 60 | 0 to 20 | 0 to 1 |
| 6 | $T_{\mathrm{Sp}}^{(\mathrm{VI})}$ | 0 to 60 | 0 to 20 | 2 to 3 |
| 7 | $T_{\mathrm{Sp}}^{(\mathrm{VII})}$ | 0 to 60 | 0 to 20 | 4 to 5 |
| 8 | $T_{\mathrm{Sp}}^{(\mathrm{VIII})}$ | 0 to 5 | 0 to 5 | 0 |
| 9 | $T_{\mathrm{Sp}}^{(\mathrm{IX})}$ | 0 to 5 | 0 to 5 | 5 |